\documentclass[12pt]{article}
\usepackage[margin=2cm]{geometry}
\usepackage{comment}
\usepackage{amsmath,amssymb,extarrows,mathtools,graphicx,subfigure,setspace}
\usepackage{cite}
\usepackage{slashed}
\usepackage{color}
\usepackage[numbers, square, sort&compress]{natbib}
\makeatother

\numberwithin{equation}{section}

\newcommand{\be}{\begin{equation}}
\newcommand{\bea}{\begin{eqnarray}}
\newcommand{\eea}{\end{eqnarray}}
\newcommand{\ba}{\begin{array}}
\newcommand{\ea}{\end{array}}
\newcommand{\ee}{\end{equation}}

\newcommand{\p}{\partial}

\begin{document}
\onehalfspacing
\noindent
\begin{titlepage}
\hfill
\vspace*{20mm}
\begin{center}
{\Large {\bf Rindler/Contracted-CFT Correspondence }\\
}

\vspace*{15mm} \vspace*{1mm} \textbf{{Reza Fareghbal$^{a,b}$,\;   Ali Naseh$^{b}$ }} 

 \vspace*{1cm}

$^a${\it  Department of Physics, 
Shahid Beheshti University, 
G.C., Evin, Tehran 19839, Iran \\ }
$^b${\it
School of Particles and Accelerators, 
 Institute for Research in Fundamental Sciences (IPM),  
 P.O. Box 19395-5531, Tehran, Iran }

\vspace*{.4cm}

{E-mails: {\tt r\_fareghbal@sbu.ac.ir, naseh@ipm.ir}}%

\vspace*{1cm}
\end{center}
\begin{abstract}
Taking the flat-space limit (zero cosmological constant limit) of
the Rindler-AdS spacetime yields the Rindler metric. According to
the proposal of Flat/contracted-CFT correspondence, the flat-space
limit on the bulk side of asymptotically AdS spacetimes corresponds to the
contraction of the conformal field theory on the boundary. We use
this proposal for the Rindler-AdS/CFT correspondence and propose a
dual theory for the Rindler spacetime, which is a contracted conformal
field theory (CCFT). We show that the two-dimensional CCFT symmetries 
exactly predict the same two-point functions that one may find by 
taking the flat-space limit of three-dimensional  Rindler-AdS 
holographic results. Using the Flat/CCFT proposal, we also calculate
the three-dimensional Rindler energy-momentum tensor. Since the near
horizon geometry of non-extreme black holes has a Rindler part, we
note  that it is plausible to find a dual CCFT at the horizon of
non-extreme black holes. By using our energy-momentum tensor, we find
the correct mass of  non-rotating BTZ and show that the
Cardy-like formula for  CCFT yields  the  Bekenstein-Hawking entropy 
of non-extreme BTZ. Our current work is the first step towards 
describing the entropy of non-extreme black holes in terms of CCFTs  
microstates which live on the horizon.

\end{abstract}
\end{titlepage}

\section{Introduction}

Recently, there has been a growing interest in studying holography
for asymptotically flat spacetimes. One particular method for studying
flat-space holography was introduced in \cite{Bagchi:2010zz,Bagchi:2012cy}
in which one starts from the known results of the AdS/CFT correspondence
and takes their flat limit. In addition to this, various aspects of the
flat-space holography were also addressed in \citep{Bagchi:2012xr,Bagchi:2012yk,
Bagchi:2013lma,Bagchi:2013bga,Afshar:2013vka,Afshar:2013bla,Gonzalez:2013oaa,
Fareghbal:2013ifa,123,Bagchilast,Grumiller:2014lna} 
by merely taking flat-space limit of the AdS counterparts.

It is a well-known fact that taking the flat-space limit of  calculations
performed in asymptotically AdS spacetimes leads to results related to the
asymptotically flat spacetimes. The flat limit of AdS is achieved  by taking
its large radius limit, which is equivalent to the zero cosmological constant
limit. At the level of action and all covariantly constant  quantities, the
flat limit of AdS clearly generates the results found in the flat case. 
However, for the metric of asymptotically AdS spacetimes, the large radius
limit must be taken carefully\footnote{An example of coordinate systems 
for asymptotically AdS spacetimes for which the flat limit is well-defined
is given in the paper \cite{Barnich:2012aw}.}. 

A related question in this regard is the interpretation of the flat limit
from the point of view of the boundary theory. It was argued in 
\cite{Bagchi:2010zz,Bagchi:2012cy} that the flat limit of asymptotically
AdS spacetimes corresponds to the contraction of coordinates of the boundary CFT.
Accordingly, asymptotically flat spacetimes have a dual description in terms
of a contracted conformal field theory (CCFT). For developing the Flat/CCFT 
correspondence one may start from AdS/CFT and take the appropriate  limit,
i.e., the  large radius limit of AdS on the bulk side and contraction of 
coordinates on the boundary. As we will discuss in  section 3, the structure
of conformal boundary dictates the coordinate which must be contracted in the CFT. 

As another check for the Flat/CCFT correspondence, one can take the flat limit
of the  Rindler-AdS/CFT correspondence. Observers with constant acceleration 
in the AdS spacetimes perceive a temperature. Moreover, they do not have access
to the full  AdS spacetime and are restricted to the two wedge-like regions
which are characterized by cosmological horizons. The coordinate system in which
these observers are at rest is called Rindler-AdS coordinates whose flat limit
is the Rindler spacetime. It was argued in \cite{Czech:2012be} that the physics
inside the two wedges of Rindler-AdS has a holographic description as  entangled
states of a pair of CFTs which live on the  boundary of Rindler-AdS wedges. The
reduced density matrix of one of these CFTs describes the spacetime inside one
of the wedges up to the horizon which should be replaced by some kind of singularity. 
Using this holographic picture, the thermodynamics of Rindler-AdS has been recovered
in \cite{Parikh:2012kg}. The temperature and entropy of Rindler-AdS are given by
using two-point function and degeneracy of states of the dual CFT. 

One can use the results of  Rindler-AdS/CFT correspondence in order to develop a
holographic dual theory for the Rindler spacetimes. Since the flat limit of 
Rindler-AdS is Rindler spacetime, it is possible to use the dictionary of Flat/CCFT
correspondence and find the CCFT dual to the Rindler. Similar to   CFTs, one can
calculate n-point functions of CCFTs (see for example \cite{represent,Akhavan:2009ns}).
In this paper, we perform the same method for the two-dimensional CCFT resulting
from the contraction
of the CFT dual to the three-dimensional Rindler-AdS and show that the two point
function of scalar operators is completely consistent with the bulk calculations.
Moreover, one can use the proposal of \cite{Fareghbal:2013ifa} for the calculation of
the stress tensor of asymptotically flat gravity. We repeat the same procedure for
the three-dimensional  Rindler spacetimes and find non-zero components of the stress
tensor which can be used in the calculation of conserved charges by using Brown
and York's method \cite{Brown:1992br}.

Furthermore, we note that two-dimensional Rindler spacetimes appear in the near
horizon geometry of non-extreme black holes. This is a property of all non-extreme
black holes with any asymptotic behaviour and in any dimension. Here, we use the
Rindler/CCFT correspondence and propose
a dual theory for the non-extreme black holes which has near horizon information. In
this view, Rindler/CCFT for non-extreme black holes is reminiscent of kerr/CFT
correspondence \cite{Guica:2008mu} for the extremal cases. For non-rotating BTZs we
check this proposal. At the first step we find the mass of non-rotating BTZ by using
our proposed stress tensor for the Rindler gravity. Finally, we find the entropy of
non-rotating BTZ by using Cardy-like formula of CCFTs which results in the 
Bekenstein-Hawking entropy. This formula has been derived in \cite{Bagchi:2012xr}
and used previously for the calculation of entropy of three-dimensional flat
cosmological solutions \cite{Cornalba:2002fi}.

The paper is organized as follows. In Section 2 we briefly review Rindler-AdS.
In Section 3 we find the CCFT dual to the three-dimensional Rindler spacetime
and calculate two point functions of scalar operators and also energy momentum
tensor of the Rindler space. Section 4 is devoted to the non-extreme/CCFT 
correspondence for which we use our proposal for finding a dual theory at
the horizon of non-extreme BTZ. In Section 5 we conclude with
some future remarks.

\section{Rindler-AdS Spacetime  }

An observer  at $r=r_0$ of three-dimensional Rindler-AdS metric, 
\begin{equation}\label{rindler-ads metric}
    ds^2=-\alpha^2r^2 d\tau^2+{dr^2\over 1+{r^2\over \ell^2}}+\left(1+{r^2\over
    \ell^2}\right)d\chi^2,
\end{equation}
perceives a constant acceleration  $a_{(3)}^2={1\over r_0^2}+{1\over \ell^2}$, where $\ell$ is AdS radius and proper time of observer is given by $\alpha r_0 \tau$. Thus the time coordinate of the metric \eqref{rindler-ads metric} is the proper time of an observer which is located at $r=r_0={1\over\alpha}$.  The temperature which these observers measure, can be given by using Rindler temperature of higher dimensional flat embeding space \cite{Deser:1997ri}. Let  $\{X^0,X^1,X^2,X^3\}$ denote the coordinates of embedding space with signature $(-,+,+,-)$. The metric \eqref{rindler-ads metric} is given by the following change of coordinates in the four-dimensional embedding space:
 \begin{eqnarray}
  \nonumber X^0 &=& r \sinh(\alpha \tau), \\
 \nonumber X^1 &=& r\cosh(\alpha \tau), \\
 \nonumber X^2 &=& \sqrt{\ell^2+r^2}\sinh\left({\chi\over \ell}\right), \\
  X^3 &=& \sqrt{\ell^2+r^2}\cosh\left({\chi\over \ell}\right).
\end{eqnarray}
Thus hyperspace $r=r_0$ is given by 
\begin{equation}
(X^1)^2-(X^0)^2= r_0^2={1\over a_{(4)}^2},
\end{equation}
where $a_{(4)}$ is four-dimensional Rindler  acceleration and the temperature is 
\begin{equation}\label{def of Temperature}
T={a_{(4)}\over 2\pi}={1\over 2\pi r_0}={\sqrt{a_{(3)}^2-{1\over\ell^2}}\over 2\pi}.
\end{equation}
Thus the Rindler-AdS observer must have acceleration greater than a critical value $1\over \ell^2$ in order to perceive a temperature. It is clear from \eqref{def of Temperature} that the temperature which an observer with proper time $\tau$ measures is 
\begin{equation}\label{Temp of rindler-ads}
T_{\text{Rindler-AdS}}={\alpha\over 2\pi}.
\end{equation}
This must coincide with the periodicity of Euclidean time in the metric \eqref{rindler-ads metric}.

The relation between Rindler-AdS coordinate and global  coordinate of  AdS$_3$ with metric
\begin{equation}\label{global ads}
ds^2=-\left(1+{\rho^2\over\ell^2}\right)dt^2+\left(1+{\rho^2\over\ell^2}\right)^{-1}d\rho^2+\rho^2d\phi^2
\end{equation}
is given by \cite{Parikh:2012kg}
\begin{eqnarray}\label{coordinate changing global}
\nonumber \rho^2&=& r^2\left[\cosh^2\left({\chi\over\ell}\right)+\sinh^2({\alpha\tau})\right]+\ell^2\sinh^2\left({\chi\over\ell}\right)\\
\nonumber \tan\phi&=& {\sqrt{r^2+\ell^2}\sinh\left({\chi\over\ell}\right)\over r\cosh(\alpha\tau)}\\
\tan\left({t\over\ell}\right)&=&{ r\sinh(\alpha\tau)\over \sqrt{r^2+\ell^2}\cosh\left({\chi\over\ell}\right)}
\end{eqnarray}
From the last equation of \eqref{coordinate changing global},  it is clear that in the limit $\tau\to\pm\infty$, we have $t\to\pm{\ell\pi\over 2}$. Thus Rindler-AdS covers a portion of global AdS which consists of two wedges. Any accelerating observer has access just to one of the wedges.  

The flat limit $(\ell\to\infty)$ of the metric \eqref{rindler-ads metric} is well-defined and results in Rindler space-time,
\begin{equation}\label{rindler metric1}
    ds^2=-\alpha^2r^2 d\tau^2+dr^2+d\chi^2.
\end{equation}


\section{From Rindler-AdS/CFT to Rindler/CCFT}

Let us start from the metric of Rindler-AdS \eqref{rindler-ads metric} and try
to find metric of its  conformal boundary. The boundary is located at $r\to\infty$.
For large and fixed $r$ we can write the metric of conformal boundary as
\begin{equation}
ds^2_{CB}={r^2\over G^2}\left(-\alpha^2 G^2 d\tau^2+{G^2\over \ell^2}d\chi^2\right),
\end{equation}
where $G$ is Newton constant. Here we intentionally used $G$ in the conformal
factor to avoid its dependence on $\ell$. Thus the  boundary is a plane with 
the metric $ds^2=-dT^2+dX^2$, where $T=\alpha G \tau$ and $X=\chi G/\ell$. 
From this identification it is clear that the flat limit on the bulk side,
i.e. $G/\ell\to 0$, corresponds to the contraction of $X$ coordinate,
i.e. $X\to\epsilon X$. Let us call the two-dimensional CFT for which the $X$
coordinate has been contracted, a contracted-CFT, or CCFT. It is argued in
the rest of this paper that holographic dual of one of the Rindler wedges is a CCFT.

The fact that in the Rindler/CCFT correspondence $X$ coordinate must be
contracted, is not in contrast with the original proposal of \cite{Bagchi:2012cy}
for the Flat/CCFT correspondence,  where time might be contracted. In fact the CFT
of \cite{Bagchi:2012cy} lives on the cylinder while in the present case it is on
the plane. In mapping from a cylinder to the plane, $X$ and $T$ change their roles.
A way for seeing this fact is using line elements \eqref{rindler-ads metric} and
\eqref{global ads}. It is easy to see that by making the changes
$\alpha\tau\to i\phi$ ,$r\to\rho$ and $\chi\to it$, the metric of Rindler-AdS
transforms to the global coordinate. These transformations are also valid for
the asymptotic Killing vectors and the generators of dual CFTs. Thus the 
contraction of time in the global case changes to the contraction of space in
the Rindler case.

\subsection{Two-point functions of scalar operators  }
For developing Rindler/CCFT, similar to Flat/CCFT correspondence, we begin with
Rindler-AdS/CFT calculations and take the appropriate limit. One of the results
of \cite{Parikh:2012kg} is that the holographic calculation for the two-point
function of two scalar operators with
conformal weights $h_1=h_2=\bar h_1=\bar h_2$, located in one of the
entangled CFTs, results in
\begin{equation}\label{twopoint}
 \langle \mathcal{O}(\chi_1,\tau_1)\mathcal{O}(\chi_2,\tau_2)\rangle={1\over\left[\cosh({\chi_1-\chi_2\over \ell})-\cosh\left(\alpha(\tau_1-\tau_2)
 \right)\right]^\Delta},
\end{equation}
where
\begin{equation}
    \Delta=h+\bar h=1+\sqrt{1+m^2\ell^2},
\end{equation}
and $m$ is the mass of corresponding scalar filed on the bulk side.

The  $\ell\to\infty$  limit (precisely ${G/\ell\to 0}$ limit )  of the two-point function
\eqref{twopoint} results in a non-zero value just for the massless case
$m=0$:

\begin{equation}\label{twopoint rindler}
 \langle \mathcal{O}(\chi_1,\tau_1)\mathcal{O}(\chi_2,\tau_2)\rangle_{rindler}={1\over\left[1-\cosh\left(\alpha(\tau_1-\tau_2)
 \right)\right]^{\Delta_{0}}},
\end{equation}
where $\Delta_0=2$.

Now we want to show that symmetry considerations lead to a similar result as
\eqref {twopoint rindler} for a CCFT. In order to find two point functions
of scalar operators in the $X$-contracted CFT or CCFT, we should find the 
symmetry generators of CCFT in the first step. A way to do this calculation
is to start from the symmetry  generators of CFT and contract them in the
manner that Flat/CCFT correspondence proposes.  

The Virasoro generators of the CFT dual to the Rindler-AdS spacetime are given by 
\begin{equation}
\mathcal{L}_n=e^{-nz}\p_z,\qquad \mathcal{\bar L}_n= e^{-n\bar z}\p_{\bar z},
\end{equation}
where $z=x+t$ and $\bar z=x-t$. 
In the Appendix we show how this representation for the conformal symmetry is 
derived\footnote{The dimensionless coordinates $x$ and $t$ are related to the
$\chi$ and $\tau$ coordinates of the bulk geometry by $t=\alpha\tau$ and $x=\chi/\ell$}.

According to the dictionary of Flat/CCFT correspondence \cite{Bagchi:2010zz,Bagchi:2012cy},
the generators of contracted  conformal algebra  are given by  
\begin{equation}\label{def of CCFT generators}
L_n=\mathcal{L}_{n}-\mathcal{\bar L}_{-n},\qquad M_n=\epsilon(\mathcal{L}_{n}+\mathcal{\bar L}_{-n}),
\end{equation}
and contracting $x\to\epsilon x$. At the $\epsilon\to 0$ limit we have
\begin{equation}\label{GCA r}
L_n=e^{-nt}(-nx\p_x+\p_t),\qquad M_n=e^{-nt}\p_x,
\end{equation}
which satisfy:
\begin{equation}\label{GCA}
[L_m,L_n]=(m-n) L_{m+n},\qquad [L_m,M_n]=(m-n) M_{m+n},\qquad  [M_m,M_n]=0.
\end{equation}
This algebra is isomorphic to the BMS$_3$ algebra \cite{Ashtekar:1996cd,Barnich:2006av}
 as the asymptotic symmetry of three dimensional asymptotically flat spacetimes. 

Since the flat limit in the bulk calculation of correlators is non-zero for the
massless case with a finite value of scaling dimension $\Delta=\Delta_0=2$, we
restrict ourselves to this particular case. In order to find two-point function
of the scalar operator, we need $\delta_{\tilde \xi} \mathcal{O}$ which is the
variation of the operator $\mathcal{O}$ under the symmetry transformation $\tilde \xi$.
We want to find this quantity for the CCFT by using the analogue transformation
in the CFT which for our case is given by
\begin{equation}
\delta_{ \xi} \mathcal{O}={\Delta_0\over 2}\left(\p_z \xi^z+\p_{\bar z} \xi^{\bar z}\right)\mathcal{O}+\left(\xi^z\p_z+\xi^{\bar z}\p_{\bar z}\right)\mathcal{O}.
\end{equation}
Using the following identities between CCFT and CFT elements,
\begin{equation}
\tilde z+\bar {\tilde z} =\lim_{\epsilon\to 0}{1\over\epsilon}(z+\bar z),\qquad \tilde z-\bar {\tilde z} =\lim_{\epsilon\to 0}(z-\bar z) ,
\end{equation}
\begin{equation}
     \xi^{\tilde z}+\xi^{\bar {\tilde z}} =\lim_{\epsilon\to 0}{1\over\epsilon}(\xi^z+ \xi^{\bar z}),\qquad  \xi^{\tilde z}-\xi^{\bar {\tilde z}}
      =\lim_{\epsilon\to 0}(\xi^z- \xi^{\bar z})
\end{equation}
\begin{equation}
\p_{\tilde z}+\p_{\bar {\tilde z}} =\lim_{\epsilon\to
0}{\epsilon}(\p_z+\p_{\bar z}),\qquad \p_{\tilde z}-\p_{\bar {\tilde
z}} =\lim_{\epsilon\to 0}(\p_z-\p_{\bar z}),
\end{equation}
where tilde shows the corresponding elements in the CCFT, one can 
check that
\begin{equation}\label{GCFT vartian}
\delta _{\tilde \xi}\mathcal{O}=\lim_{\epsilon\to 0}\delta_{ \xi}
\mathcal{O}={\Delta_0\over 2}\left(\p_{\tilde z} \xi^{\tilde
z}+\p_{\bar {\tilde z}} \xi^{\bar {\tilde
z}}\right)\mathcal{O}+\left(\xi^{\tilde z}\p_{\tilde z}+\xi^{\bar
{\tilde z}}\p_{\bar {\tilde z}}\right)\mathcal{O}.
\end{equation}

Lets denote the two-point function of scalar operator in the CCFT by
$G_2(t_1,x_1;t_2,x_2)=\langle\mathcal{O}(t_1,x_1)\mathcal{O}(t_2,x_2)\rangle$. 
This must be invariant under the $M_n$ and $L_n$ for $n=0,\pm1$. Using
\eqref{GCA r} and \eqref{GCFT vartian}, we find that
\begin{equation}
    \left(e^{-n t_1}\p_{x_1}+e^{-n t_2}\p_{x_2}\right) G_{2}(t_1,x_1;t_2,x_2)=0,
\end{equation}
\begin{equation}
    \left[-n\Delta_0(e^{-n t_1}+e^{-n t_2})+e^{-n t_1}(\p_{t_1}-n x_1 \p_{x_1})+e^{-n t_2}(\p_{t_2}-n x_2
    \p_{x_2})\right] G_{2}(t_1,x_1;t_2,x_2)=0.
\end{equation}
 These equations determine the two-point function as
\begin{equation}\label{two point in GCFT}
    G_2(t_1,x_1;t_2,x_2)={C\over \left(\sinh{t_1-t_2\over
    2}\right)^{2\Delta_0}},
\end{equation}
which is the same as \eqref{twopoint rindler}.

\subsection{A Stress Tensor for Rindler Spacetime}
Using Rindler/CCFT correspondence, we can define a stress tensor for the Rindler spacetime.
The connection is similar to the generic rule of AdS/CFT, i.e., the stress tensor of gravity
is given by one pint function of energy-momentum tensor of dual theory. Since in developing
Flat/CCFT correspondence we find all quantities by the flat limit of AdS/CFT calculations,
we apply the same procedure for the stress tensor as well. In fact the method of calculation
of the flat-space energy-momentum tensor has been introduced earlier in the paper 
\cite{Fareghbal:2013ifa}\footnote{The holographic renormalization approach for the
flat space holography has been also studied in \cite{555} and \cite{666}, but non of them
addressed the connection to contracted CFT.}. The stress tensor of flat gravity is given
by one point functions of the contracted CFT resulting from contracting the one point 
functions of the original CFT. In this section we use the same method for finding the stress
tensor of three-dimensional Rindler spacetimes. The starting point is Rindler-AdS metric
\eqref{rindler-ads metric} and the definition of Brown and York's quasi-local energy momentum
tensor \cite{Brown:1992br},
\begin{equation}\label{def of EM}
 T^{\mu\nu} = \frac{2}{\sqrt{-\gamma}}\frac{\delta S}{\delta\gamma_{\mu\nu}},
 \end{equation} 
where $S= S_{grav}(\gamma_{\mu\nu})$ is the gravitational action viewed as
a functional of boundary metric $\gamma_{\mu\nu}$,
\begin{equation}\label{AdS action}
S = \frac{1}{16\pi G}\int _{\mathcal{M}} d^{3}x \sqrt{-g}
\left( R- \frac{2}{\ell^{2}}\right) -\frac{1}{8\pi G} \int_{\partial\mathcal{M}}
d^{2}x \sqrt{-\gamma} \;\mathcal{K}-\frac{1}{8\pi G\ell}\int _{\partial\mathcal{M}}d^2x\sqrt{-\gamma}.
\end{equation}
The second term is  Gibbons-Hawking term and the last one  is the
counterterm action that  must be added in order to obtain a finite stress tensor, and
$\mathcal{K}$ is trace of the extrinsic curvature of the boundary
\begin{eqnarray}
\mathcal{K} &=& \gamma^{\mu\nu}\mathcal{K}_{\mu\nu} \;=\;
\gamma^{\mu\nu}\gamma_{\mu} ^{\rho}\;\nabla_{\rho}n_{\nu},
\end{eqnarray}
where $\gamma_{\mu\nu} = g_{\mu\nu}-n_{\mu}n_{\nu}$ 
and $n_{\nu}$ is the outward pointing normal vector to the boundary $\partial\mathcal{M}$.
\eqref{def of EM} and \eqref{AdS action} result in
\begin{equation}\label{stress tensor formula}
T_{\mu\nu} = -\frac{1}{8\pi G} \left( \mathcal{K}_{\mu\nu}
-\mathcal{K}\gamma_{\mu\nu} +{\gamma_{\mu\nu}\over\ell}\right).
\end{equation}

At the boundary of Rindler-AdS \eqref{rindler-ads metric} i.e. $r\to\infty$, the non-zero components of $T_{\mu\nu}$ are
 \begin{equation}\label{stress tensor rindler-ads}
T_{\tau\tau}={\alpha^2\ell\over 16\pi G},\qquad T_{\chi\chi}={1\over 16\pi G\ell}.
\end{equation}
It is clear that the flat limit $G/\ell\to 0$ is not well-defined for
\eqref{stress tensor rindler-ads}. In order to define stress tensor of
Rindler spacetime ,$\tilde T_{\mu\nu}$, by taking the flat limit of
the stress tensor of Rindler-AdS ,$T_{\mu\nu}$, we use the proposal of
\cite{Fareghbal:2013ifa} and write
\begin{eqnarray}\label{def of rindler stress}
\nonumber\tilde T_{++}+ \tilde T_{--}&=&\lim_{{G\over\ell}\to 0}\left[{G\over\ell}(T_{++}+T_{--})\right],\\
\tilde T_{++}- \tilde T_{--}&=&\lim_{{G\over\ell}\to 0}\left[T_{++}-T_{--}\right],
\end{eqnarray}
where the light-cone coordinates in the Rindler-AdS  are given by $x^{\pm}={G\over\ell}\chi\pm\alpha G \tau$,
while in Rindler spacetime we define the light-cone coordinates as $\tilde x^{\pm}=\chi\pm\alpha G \tau$ with
$\tau$ and $\chi$ being coordinates of \eqref{rindler metric1}. The $\tilde T_{+-}$ component like $T_{+-}$
is assumed to be zero. The reader can easily check that with the definition \eqref{def of rindler stress},
$\tilde T_{++}$ and $\tilde T_{--}$ are both finite, and finally for the Rindler spacetime we have
\begin{equation}\label{stress tensor rindler}
\tilde T_{\tau\tau}={\alpha^2\over 16\pi },\qquad \tilde T_{\chi\chi}={1\over 16\pi G^2}.
\end{equation}
Since the dual CCFT of Rindler is constructed by contracting space-like coordinate i.e. $,x\to\epsilon x$,
it is natural to expect that its momentum operator is found from the original CFT by a similar scaling.
The first line of \eqref{def of rindler stress} is the corresponding equation for the scaling of momentum
from the bulk point of view. The second equation of \eqref{def of rindler stress} shows that Hamiltonian
of CCFT is the same as that of the original CFT. This is justified by the fact that our contraction does
not affect the time coordinate.

The  point which we used implicitly in the definition \eqref{def of rindler stress} of the stress tensor
for Rindler spacetime (but will be explicitly used in the rest of this section) is that we assume that 
similar to the original CFT, CCFT lives on a  plane. The metric of this plane is similar to the CFT
case but with the difference that its $x$ coordinate is the contracted one. This fact is clear in the
definition of the light-cone coordinates $x^\pm$ and $\tilde x^\pm$ for Rindler-AdS and Rindler spacetimes,
respectively, where the limit $G/\ell\to0$ in the bulk corresponds to the limit $\epsilon\to 0$ on the boundary. 
In summary, the CCFT dual to the Rindler spacetime 
\begin{equation}\label{rindler metric}
    ds^2=-\alpha^2r^2 d\tau^2+dr^2+d\chi^2,
\end{equation}
lives on a plane with metric
\begin{equation}\label{metric of boundary theory for Rindler}
ds^2_{\p_M}=-\alpha^2 G^2 d\tau^2+d\chi^2.
\end{equation}

\section{Non-extreme black holes/CCFT correspondence}
The energy momentum of previous section for Rindler spacetime can be used
in the computation of conserved charges of non-extreme BTZ. If one takes
the near horizon limit of non-extreme black holes with any asymptotic behaviour,
the resulting geometry has a  Rindler part. Let us study the simplest case which
is non-rotating BTZ with the metric 
\begin{equation}
ds^2=-f(\rho)d\tau^2+f(\rho)^{-1}d\rho^2+\rho^2d\phi^2,
\end{equation}
where $f(\rho)={\rho^2\over\ell^2}-8GM$ and $M$ is the mass of the black hole. If
we denote the outer horizon by $\rho_h$, which is given by $\rho_h=\ell\sqrt{8GM}$, 
and define a new coordinate $y=\rho-\rho_h$, in the near horizon characterized by
$y\ll \rho_h$, we find 
\begin{equation}
ds_{NH}^2=-f'(\rho_h) y d\tau^2+(f'(\rho_h) y)^{-1}dy^2+\rho_h^2 d\phi^2.
\end{equation}
Defining a new coordinate $r$ by $dr=dy/\sqrt{f'(\rho_h) y}$ results in
\begin{equation}\label{NHNExtreme}
ds^2_{NH}=-{f'^2(\rho_h)\over 4} r^2 d\tau^2+dr^2+\rho_h^2 d\phi^2.
\end{equation}
Comparing \eqref{NHNExtreme} with \eqref{rindler metric} shows that the near
horizon metric is Rindler with $\alpha=f'(\rho_h)/2=\sqrt{8GM}/\ell$ and a compact $\chi$ coordinate,
given by $\chi=\rho_h\phi$ \footnote{$\ell$ in the parameter $\alpha$ and also $\rho_h$  is not radius
of AdS in the metric of Rindler-AdS. Thus taking the limit $\ell\to\infty$ from the calculations of
 Rindler-AdS does not affect $\alpha$ and $\rho_h$.}.

Using  \eqref{stress tensor rindler}, we find the non-zero component of the stress tensor as 
\begin{equation}\label{stress tensor rindler of nonextreme}
\tilde T_{\tau\tau}={\alpha^2\over 16\pi }={G M \over 2\pi\ell^2},\qquad \tilde T_{\phi\phi}={\rho_h^2\over 16\pi G^2}={\ell^2 M\over 2\pi G}.
\end{equation}

Let us use \eqref{stress tensor rindler of nonextreme} and compute the conserved charge of killing vector $\xi=\p_\tau$.
If we use the metric
\begin{equation}\label{metric of boundary theory for Rindler nonextrem}
ds^2_{\p_M}=-\alpha^2 G^2 d\tau^2+\rho_h^2 d\phi^2,
\end{equation}
in the Brown and York's formula
\begin{equation}
 Q_\xi=\int_\Sigma\, d\phi \sqrt{\sigma} v^\mu \xi^\nu \tilde T_{\mu\nu},
 \end{equation} 
where $\Sigma$ is the spacelike surface of $\partial M$ ($\tau=$ constant surface),
 $\sigma_{ab}$ is metric of $\Sigma$ i.e. $\sigma_{ab}dx^a dx^b= \rho_h^2 d\phi^2$ and $v^\mu$ is the
unit time-like vector normal to $\Sigma$, we find
\begin{equation}
Q_{\p_\tau}={\alpha\rho_h\over 8G}=M.
\end{equation}
Thus our stress tensor results in the mass of the non-rotating BTZ as the charge of killing vector $\xi=\p_\tau$,
where $\tau$ is dimension-full time of near horizon geometry. This is a non-trivial check for our proposed
 energy-momentum tensor of Rindler spacetime. 

\subsection{Entropy of non-extreme BTZ as degeneracy of CCFT states}
Our discussion so far, proposes a dual two-dimensional CCFT for the three-dimensional
Rindler spacetime. Moreover, since the near horizon geometry of non-extreme black holes
has a Rindler part, it is plausible to propose a dual description for the non-extreme
black holes at the horizon. This dual theory is a CCFT, which has contracted conformal
symmetry. In two-dimensions this symmetry is given by \eqref{GCA}. If we start from two
copies of conformal algebra with central charges $c$ and $\bar c$ and define generators
of CCFT as \eqref{def of CCFT generators}, the $[L_n,L_m]$ and $[L_n,M_m]$ commutators
will have central extensions. Let us denote the central charge of $[L_n,L_m]$ by $c_L$
and central charge of $[L_n,M_m]$ by $c_M$. They are given by taking the following limit
of the central charges of the original CFT:
\begin{eqnarray}
\nonumber c_L &=& \lim_{\epsilon\to 0} \frac{1}{12}(c-\bar c),\\
c_M &=& \lim_{\epsilon\to 0} \frac{\epsilon}{12}(c+\bar c).
\end{eqnarray}

According to the Flat/CCFT proposal, the $\epsilon\to 0$ limit on the boundary theory
corresponds to $G/\ell\to 0$ in the bulk. If we use Brown and Henneaux's central charges
$c=\bar c=3\ell/2G$ \cite{Brown:1986nw}, the central charge $c_L$ will be zero but $c_M=1/4$.  

The degeneracy of states $|h_L,h_M\rangle$ which are eigenstates of $L_0$ and $M_0$ with
eigenvalues $h_L$ and $h_M$, can be given by a Cardy like formula for the CCFT. This 
formula has been given in \cite{Bagchi:2012xr} for large charges  as 
\begin{equation}\label{Cardy-like}
 S =  \ln d(h_L, h_M) =  2\pi\bigg( c_{L}
\sqrt{\frac{h_M}{2c_{M}}} + h_L \sqrt{\frac{c_{M}}{2h_M}}
\bigg),
 \end{equation}
where $(h_L,h_M)$ are given in terms of conformal weights $(h,\bar h)$ of the original CFT as
\begin{equation}\label{def of gca weights}
h_L=\lim_{\epsilon\to 0}(h-\bar h),\qquad h_M=\lim_{\epsilon\to 0}\epsilon(h+\bar h),
\end{equation}  
  
If our proposal for the dual CCFT at the horizon of non-extreme black holes is correct,
the entropy of non-extreme BTZ must be given by such a formula in terms of degeneracy
of states of CCFT. In the following we will show this fact for the entropy of non-rotating BTZ.

The first step is the calculation of $h_M$ and $h_L$. They are given by \eqref{def of gca weights}
from conformal weights $(h,\bar h)$ of original CFT. The original CFT is dual to  Rindler-AdS 
spacetimes for which the flat limit results in near horizon geometry of BTZ. Thus the metric of
Rindler-AdS is given by \eqref{rindler-ads metric} but its $\chi$ coordinate should be compactified
as $\chi=\rho_h\phi$, where $\rho_h$ is radius of horizon of BTZ and it is just a parameter in this
stage; the $\phi$ coordinate takes values in the interval $[0,2\pi]$. Now the entropy of the 
cosmological horizon of Rindler-AdS located at $r=0$ is  finite and is given by $S=\pi \rho_h/2G$.
Using the Rindler-AdS/CFT correspondence, this entropy must be given by Cardy formula in terms of
degeneracy of states of CFT. Let us write Cardy formula in the following form 
\begin{equation}
S={\pi^2\over 3}(c T_R+\bar c T_L),
\end{equation}
where right and left temperatures of CFT are given by 
\begin{equation}
\dfrac{1}{T_R}=\pi\sqrt{c\over 6 h},\qquad \dfrac{1}{T_L}=\pi\sqrt{\bar c\over 6 \bar h}.
\end{equation}
Moreover, $T_L$ and $T_R$ are related to the temperature $T=\alpha/2\pi$ of Rindler-AdS as 
\begin{equation}
\frac{1}{T}=\frac{\ell}{2}\left(\frac{1}{T_R}+\frac{1}{T_L}\right).
\end{equation}
 Thus we can calculate $h$ and $\bar h$. The final result is
 \begin{eqnarray}\label{values of h bh}
 \nonumber h&=& {\rho_h^2\over16\ell G}\left(2-{\alpha\ell^2\over\rho_h}+2\sqrt{1-{\alpha\ell^2\over\rho_h}}\right),\\
     \bar h&=& {\rho_h^2\over16\ell G}\left(2-{\alpha\ell^2\over\rho_h}-2\sqrt{1-{\alpha\ell^2\over\rho_h}}\right)
 \end{eqnarray}
 
Now we can calculate $(h_L,h_M)$ of CCFT dual to Rindler spacetime. They are
given by \eqref{def of gca weights} where $\epsilon$ must be identified with
$G/\ell$. Using \eqref{values of h bh}, the $G/\ell\to 0$  is well-defined and
results in:
\begin{equation}\label{final values of hl hm}
h_L={\rho_h^2\over 4G}\sqrt{-{\alpha\over\rho_h}},\qquad h_M=-{\alpha \rho_h\over 8}.
\end{equation}
Putting everything in the Cardy-like formula \eqref{Cardy-like} gives 
\begin{equation}
S={\pi\rho_h\over 2 G},
\end{equation}
which is exactly the Bekenstein-Hawking  entropy of non-extremal BTZ.

\section{Conclusion}
Our calculations in this paper uncover the existence of a two-dimensional dual
CCFT for gravity in three-dimensional Rindler spacetimes. There are several
points related to this duality which we did not address in this paper. The
hypersurface on which CCFT lives, i.e., the metric \eqref{metric of boundary theory for Rindler}
is constructed by an anisotropic scaling of Rindler space. It seems that this kind
of anisotropic scaling is necessary for studying flat-space holography. 
Moreover, for the current case of Rindler holography,  the dual theory is
at spacial infinity rather than Null infinity of flat spacetimes. This point
does not in contrast with previously known results about flat-space holography
but needs more study\footnote{ The BMS symmetry also exist at the spacial
infinity of asymptotically flat spacetimes\cite{Ashtekar:1978zz}.}.

The independence of correlation function \eqref{two point in GCFT} from space
coordinates is another interesting property of Rindler-space holography. These
correlators may reveal some properties of CCFTs dual to asymptotically flat
spacetimes. 

The non-extreme black holes/CCFT correspondence which we propose in this paper
is the first step in this direction and needs careful studies in higher dimensions.
Since there is no decoupling between near-horizon region and rest of space for
the non-extreme  black holes, the information which the horizon CCFT may provide
is problematic. The main question is how much information this dual theory has
about the inside region of horizon, its hairs and entropy of black hole. Another
point which relations \eqref{final values of hl hm} reveal, is the unusual 
values of $h_L$ and $h_M$. One of them is negative and another is imaginary, but
in the Cardy-like formula they compensate each other and the final entropy is
well-defined. The relation between CCFTs  and the proposals of  \cite{Carlip:1998wz,Solodukhin:1998tc}
for the dual theory of horizons which has a chiral Virasoro  symmetry is another
interesting question in this regard\footnote{ A comment on the relation of this
chiral Virasoro and equipartition of energy is given  in section 4.3 of review \cite{Simon:2011zza}.}.

All the above points are signs of non-triviality of CCFTs dual to flat spacetimes and may uncover
new facts about  gravity in the flat spacetimes. We hope to address the above points in our future works.

\section*{Acknowledgement}
The authors  would like to  thank M. Alishahiha, D. Allahbakhshi, A. Bagchi, D. Grumiller, A. Hosseiny, A. E. Mosaffa, M. Safari and J. Simon for useful comments and discussions. We also especially thank D. Kaviani for his comments on the revised version.

\section*{Appendix}
\setcounter{equation}{0}
\subsection*{Representation of CFT generators dual to the Rindler-AdS }
Symmetry generators can be given by starting from Killing vectors of Rindler-AdS \eqref{rindler-ads metric}, restricting them to the $(\tau,\chi)$ components and   taking $r\to\infty$ limit. This could result in the global part of conformal algebra and one can extend them to find full symmetry. Using generators of 4 dimensional flat embedding space, we can find Killing vectors of Rindler-AdS as
\begin{equation}
J_{01}=-\dfrac{1}{\alpha} \p_\tau
\end{equation}
\begin{equation}
J_{02}=-{r\ell\over\sqrt{r^2+\ell^2}}\sinh(\alpha\tau)\cosh\left(\chi\over\ell\right)\p_\chi-\sqrt{r^2+\ell^2}\sinh\left(\chi\over\ell\right)\left[{\cosh(\alpha\tau)\over\alpha r}\p_\tau-\sinh(\alpha\tau)\p_r\right]
\end{equation}
\begin{equation}
J_{03}={r\ell\over\sqrt{r^2+\ell^2}}\sinh(\alpha\tau)\sinh\left(\chi\over\ell\right)\p_\chi+\sqrt{r^2+\ell^2}\cosh\left(\chi\over\ell\right)\left[{\cosh(\alpha\tau)\over\alpha r}\p_\tau-\sinh(\alpha\tau)\p_r\right]
\end{equation}
\begin{equation}
J_{12}=-{r\ell\over\sqrt{r^2+\ell^2}}\cosh(\alpha\tau)\cosh\left(\chi\over\ell\right)\p_\chi+\sqrt{r^2+\ell^2}\sinh\left(\chi\over\ell\right)\left[{\sinh(\alpha\tau)\over\alpha r}\p_\tau-\cosh(\alpha\tau)\p_r\right]
\end{equation}
\begin{equation}
J_{13}=-{r\ell\over\sqrt{r^2+\ell^2}}\cosh(\alpha\tau)\sinh\left(\chi\over\ell\right)\p_\chi-\sqrt{r^2+\ell^2}\cosh\left(\chi\over\ell\right)\left[{\sinh(\alpha\tau)\over\alpha r}\p_\tau-\cosh(\alpha\tau)\p_r\right]
\end{equation}
\begin{equation}
J_{23}=\ell\p_\chi
\end{equation}
Taking $r\to\infty$ and defining $t=\alpha\tau$ and $x=\chi/\ell$ and restricting to the $x$ and $t$ components gives the generators of  global part of conformal algebra as
\begin{equation}
j_{01}=-\p_t
\end{equation}
\begin{equation}
j_{02}=-\sinh t \cosh x \p_x-\sinh x\cosh t \p_t
\end{equation}
\begin{equation}
j_{03}=\sinh t \sinh x \p_x+\cosh x\cosh t \p_t
\end{equation}
\begin{equation}
j_{12}=\cosh t \cosh x \p_x+\sinh x\sinh t \p_t
\end{equation}
\begin{equation}
j_{13}=-\cosh t \sinh x \p_x-\cosh x\sinh t \p_t
\end{equation}
\begin{equation}
j_{23}=\p_x
\end{equation}
The   $SL(2,R)\times SL(2,R)$ generators are given by
\begin{equation}
\mathcal{L}_0=\dfrac{1}{2}(j_{23}-j_{01}),\qquad \mathcal{L}_{+1}=\frac{1}{2}(j_{02}+j_{03}+j_{12}+j_{13}),\qquad \mathcal{L}_{-1}=-\frac{1}{2}(j_{02}-j_{03}-j_{12}+j_{13})
\end{equation}
 \begin{equation}
\mathcal{\bar L}_0=\dfrac{1}{2}(j_{23}+j_{01}),\qquad \mathcal{\bar L}_{-1}=\frac{1}{2}(j_{02}-j_{03}+j_{12}-j_{13}),\qquad \mathcal{\bar L}_{1}=-\frac{1}{2}(j_{02}+j_{03}-j_{12}-j_{13})
\end{equation}
which can be written as
\begin{equation}
\mathcal{L}_n=e^{-nz}\p_z,\qquad \mathcal{\bar L}_n= e^{-n\bar z}\p_{\bar z}
\end{equation}
where $z=x+t$ and $\bar z=x-t$.


\begin{thebibliography}{}




\bibitem{Bagchi:2010zz}
  A.~Bagchi,
  ``Correspondence between Asymptotically Flat Spacetimes and Nonrelativistic Conformal Field Theories,''
  Phys.\ Rev.\ Lett.\  {\bf 105}, 171601 (2010).

  A.~Bagchi,
  ``The BMS/GCA correspondence,''
  arXiv:1006.3354 [hep-th].

\bibitem{Bagchi:2012cy}
  A.~Bagchi and R.~Fareghbal,
  ``BMS/GCA Redux: Towards Flatspace Holography from Non-Relativistic Symmetries,''
  JHEP {\bf 1210}, 092 (2012)
  [arXiv:1203.5795 [hep-th]].


\bibitem{Bagchi:2012xr} 
  A.~Bagchi, S.~Detournay, R.~Fareghbal and J.~Simon,
  ``Holography of 3d Flat Cosmological Horizons,''
  Phys.\ Rev.\ Lett.\  {\bf 110}, 141302 (2013)
  [arXiv:1208.4372 [hep-th]].


\bibitem{Bagchi:2012yk} 
  A.~Bagchi, S.~Detournay and D.~Grumiller,
  ``Flat-Space Chiral Gravity,''
  Phys.\ Rev.\ Lett.\  {\bf 109}, 151301 (2012)
  [arXiv:1208.1658 [hep-th]]. 
 



\bibitem{Bagchi:2013lma} 
  A.~Bagchi, S.~Detournay, D.~Grumiller and J.~Simon,
  ``Cosmic evolution from phase transition of 3-dimensional flat space,''
  Phys.\ Rev.\ Lett.\  {\bf 111}, 181301 (2013)
  [arXiv:1305.2919 [hep-th]].
 
 
 
 

%
 
 
\bibitem{Bagchi:2013bga} 
  A.~Bagchi,
  ``Tensionless Strings and Galilean Conformal Algebra,''
  JHEP {\bf 1305}, 141 (2013)
  [arXiv:1303.0291 [hep-th]].







\bibitem{Afshar:2013vka}
  H.~Afshar, A.~Bagchi, R.~Fareghbal, D.~Grumiller and J.~Rosseel,
  ``Higher spin theory in 3-dimensional flat space,''
  Phys.\ Rev.\ Lett.\  {\bf 111} (2013) 121603
  [arXiv:1307.4768 [hep-th]].











\bibitem{Afshar:2013bla} 
  H.~R.~Afshar,
  ``Flat/AdS boundary conditions in three dimensional conformal gravity,''
  JHEP {\bf 1310}, 027 (2013)
  [arXiv:1307.4855 [hep-th]].
  
  
  
\bibitem{Gonzalez:2013oaa} 
  H.~A.~Gonzalez, J.~Matulich, M.~Pino and R.~Troncoso,
  ``Asymptotically flat spacetimes in three-dimensional higher spin gravity,''
  JHEP {\bf 1309}, 016 (2013)
  [arXiv:1307.5651 [hep-th]].
  
  
 


















\bibitem{Fareghbal:2013ifa} 
  R.~Fareghbal and A.~Naseh,
  ``Flat-Space Energy-Momentum Tensor from BMS/GCA Correspondence,''
  JHEP {\bf 1403}, 005 (2014)
  [arXiv:1312.2109 [hep-th]].
  
  
  
  





























\bibitem{123} 
  C.~Krishnan, A.~Raju and S.~Roy,
  ``A Grassmann Path From AdS$_3$ to Flat Space,''
  arXiv:1312.2941 [hep-th].

\bibitem{Bagchilast} 
  A.~Bagchi and R.~Basu,
  ``3D Flat Holography: Entropy and Logarithmic Corrections,''
  arXiv:1312.5748 [hep-th].





\bibitem{Grumiller:2014lna} 
  D.~Grumiller, M.~Riegler and J.~Rosseel,
  ``Unitarity in three-dimensional flat space higher spin theories,''
  arXiv:1403.5297 [hep-th].




\bibitem{Barnich:2012aw} 
  G.~Barnich, A.~Gomberoff and H.~A.~Gonzalez,
  ``The Flat limit of three dimensional asymptotically anti-de Sitter spacetimes,''
  Phys.\ Rev.\ D {\bf 86}, 024020 (2012)
  [arXiv:1204.3288 [gr-qc]].

\bibitem{Czech:2012be}
  B.~Czech, J.~L.~Karczmarek, F.~Nogueira and M.~Van Raamsdonk,
  ``Rindler Quantum Gravity,''
  Class.\ Quant.\ Grav.\  {\bf 29}, 235025 (2012)
  [arXiv:1206.1323 [hep-th]].

\bibitem{Parikh:2012kg}
  M.~Parikh and P.~Samantray,
  ``Rindler-AdS/CFT,''
  arXiv:1211.7370 [hep-th].
  
  
  

\bibitem{represent}
  A.~Bagchi and I.~Mandal,
  ``On Representations and Correlation Functions of Galilean Conformal Algebras,''
  Phys.\ Lett.\ B {\bf 675}, 393 (2009)
  [arXiv:0903.4524 [hep-th]]
 
 
\bibitem{Akhavan:2009ns} 
  A.~Akhavan, M.~Alishahiha, A.~Davody and A.~Vahedi,
  ``Non-relativistic CFT and Semi-classical Strings,''
  JHEP {\bf 0903}, 053 (2009)
  [arXiv:0811.3067 [hep-th]].
 


\bibitem{Brown:1992br} 
  J.~D.~Brown and J.~W.~York, Jr.,
  ``Quasilocal energy and conserved charges derived from the gravitational action,''
  Phys.\ Rev.\ D {\bf 47}, 1407 (1993)
  [gr-qc/9209012].



\bibitem{Guica:2008mu} 
  M.~Guica, T.~Hartman, W.~Song and A.~Strominger,
  ``The Kerr/CFT Correspondence,''
  Phys.\ Rev.\ D {\bf 80}, 124008 (2009)
  [arXiv:0809.4266 [hep-th]].




\bibitem{Cornalba:2002fi} 
  L.~Cornalba and M.~S.~Costa,
  ``A New cosmological scenario in string theory,''
  Phys.\ Rev.\ D {\bf 66}, 066001 (2002)
  [hep-th/0203031]. 












\bibitem{Deser:1997ri} 
  S.~Deser and O.~Levin,
  ``Accelerated detectors and temperature in (anti)-de Sitter spaces,''
  Class.\ Quant.\ Grav.\  {\bf 14}, L163 (1997)
  [gr-qc/9706018].\\
  S.~Deser and O.~Levin,
  ``Equivalence of Hawking and Unruh temperatures through flat space embeddings,''
  Class.\ Quant.\ Grav.\  {\bf 15}, L85 (1998)
  [hep-th/9806223].\\
   S.~Deser and O.~Levin,
  ``Mapping Hawking into Unruh thermal properties,''
  Phys.\ Rev.\ D {\bf 59}, 064004 (1999)
  [hep-th/9809159].\\
  T.~Jacobson,
  ``Comment on `Accelerated detectors and temperature in anti-de Sitter spaces',''
  Class.\ Quant.\ Grav.\  {\bf 15}, 251 (1998)
  [gr-qc/9709048].

\bibitem{Ashtekar:1996cd} 
  A.~Ashtekar, J.~Bicak and B.~G.~Schmidt,
  ``Asymptotic structure of symmetry reduced general relativity,''
  Phys.\ Rev.\ D {\bf 55}, 669 (1997)
  [gr-qc/9608042].
  

\bibitem{Barnich:2006av} 
  G.~Barnich and G.~Compere,
  ``Classical central extension for asymptotic symmetries at null infinity in three spacetime dimensions,''
  Class.\ Quant.\ Grav.\  {\bf 24}, F15 (2007)
  [gr-qc/0610130].




\bibitem{555} 
  R.~N.~C.~Costa,
  ``Aspects of the zero $\Lambda$ limit in the AdS/CFT correspondence,''
  arXiv:1311.7339 [hep-th].



\bibitem{666} 
  S.~Detournay, D.~Grumiller, F.~Scholler and J.~Simon,
  ``Variational principle and 1-point functions in 3-dimensional flat space Einstein gravity,''
  arXiv:1402.3687 [hep-th].



 
\bibitem{Brown:1986nw} 
  J.~D.~Brown and M.~Henneaux,
  ``Central Charges in the Canonical Realization of Asymptotic Symmetries: An Example from Three-Dimensional Gravity,''
  Commun.\ Math.\ Phys.\  {\bf 104}, 207 (1986). 



\bibitem{Ashtekar:1978zz} 
  A.~Ashtekar and R.~O.~Hansen,
  ``A unified treatment of null and spatial infinity in general relativity. I - Universal structure, asymptotic symmetries, and conserved quantities at spatial infinity,''
  J.\ Math.\ Phys.\  {\bf 19}, 1542 (1978).




\bibitem{Carlip:1998wz}
  S.~Carlip,
  ``Black hole entropy from conformal field theory in any dimension,''
  Phys.\ Rev.\ Lett.\  {\bf 82} (1999) 2828
  [hep-th/9812013].


\bibitem{Solodukhin:1998tc}
  S.~N.~Solodukhin,
  ``Conformal description of horizon's states,''
  Phys.\ Lett.\ B {\bf 454} (1999) 213
  [hep-th/9812056].

\bibitem{Simon:2011zza}
  J.~Simon,
  ``Extremal black holes, holography and coarse graining,''
  Int.\ J.\ Mod.\ Phys.\ A {\bf 26} (2011) 1903
  [arXiv:1106.0116 [hep-th]].






\end{thebibliography}
\end{document}